\let\pdfoutput\defined

\documentclass{PoS}
\usepackage{amsmath, amsfonts, bbm, dsfont, mathrsfs,amssymb}
\usepackage{dcolumn}
\usepackage{subfigure}

\title{%
\vspace*{-1cm}
\begin{minipage}{\textwidth}
\begin{flushright}
\texttt{\footnotesize
PoS(LAT2007)120\\%
BNL-HET-07/17\\%
}
\end{flushright}
\end{minipage}\\[15pt]
Chiral Limit and Light Quark Masses\\ in 2+1 Flavor Domain Wall QCD%
}

\ShortTitle{Chiral Limit and Light Quark Masses in 2+1 Flavor Domain Wall QCD}

\author{RBC and UKQCD Collaborations} 

\author{Meifeng Lin\thanks{Speaker, presentation:\textit{Probing the
Chiral Limit in 2+1 Flavor Domain Wall Fermion QCD}.} \thanks{Current
address: Center for Theoretical Physics, Massachusetts Institue of
Technology, Cambridge, MA 02139, USA}\\
        Physics Department, Columbia University, New York, NY 10027, USA\\
        E-mail: \email{mflin@phys.columbia.edu}}

\author{Enno E. Scholz\thanks{Speaker, presentation:\textit{Quark Mass Determination from 2+1 Flavor Domain Wall Fermion Simulations }.}\\
        Physics Department, Brookhaven National Laboratory, Upton, NY 11973, USA\\
        E-mail: \email{scholzee@quark.phy.bnl.gov}}

\abstract{%
We present results for meson masses and decay constants measured on $24^3\times64$ lattices using the domain wall fermion formulation with an extension of the fifth dimension of $L_s=16$ for $N_f=2+1$ dynamical quark flavors. The lightest dynamical meson mass in our set-up is around $331\,{\rm MeV}$, while partially quenched mesons reach masses as low as $250\,{\rm MeV}$. The applicability of $\SU(3)\times\SU(3)$ and $\SU(2)\times\SU(2)$ (partially quenched) chiral perturbation theory will be compared and we quote values for the low-energy constants from both approaches. We will extract the average light quark and strange quark masses and use a non-perturbative renormalization technique (RI/MOM) to quote their physical values. The pion and kaon decay constants are determined at those values from our chiral fits and their ratio is used to obtain the CKM-matrix element $|V_{us}|$. The results presented here include statistical errors only.%
}

\FullConference{The XXV International Symposium on Lattice Field Theory\\
		 July 30 -- August 4 2007\\
		 Regensburg, Germany}

\newcommand{\mres}{m_{\rm res}}

\def\nicefrac#1#2{\leavevmode\kern.1em\raise.5ex\hbox{\the\scriptfont0 #1}\kern-.1em/\kern-.15em\lower.25ex\hbox{\the\scriptfont0 #2}}

\newcommand{\SU}{{\rm SU}}

\newcommand{\sci}[1]{\cdot10^{#1}}

\newcolumntype{C}{>{$}c<{$}}
\newcolumntype{R}{>{$}r<{$}}
\newcolumntype{L}{>{$}l<{$}}

\newlength{\closercaption}
\newlength{\afterTable}
\newlength{\afterFigure}
\begin{document}

$\phantom{\speaker{M.F.~Lin and E.E.~Scholz}}$
\vspace*{-1cm}

Lattice QCD allows us to study QCD phenomenology from first principles
by using Monte Carlo techniques. Recent developments in both the
computer technology and numerical algorithms have made possible
lattice simulations with the correct number of fermion flavors in the
vacuum polarization, which are essential for establishing direct
connections between lattice simulations and the underlying low-energy
QCD.  However, the computational cost increases dramatically as one
decreases the quark masses in the simulations towards the chiral
limit. As such, current lattice simulations still work with quark
masses heavier than their physical values, and extrapolations are
necessary to obtain meaningful physical results from the simulations
with heavy quark masses. 

Chiral perturbation theory ($\chi$PT) is a low-energy effective theory
which connects physical observables to quark masses in explicit
functional  forms, and is a useful tool to guide the extrapolations
for the  lattice QCD simulations. Since it is based on the approximate
chiral symmetry of QCD,  it is important to have a chiral fermion
formulation  on the lattice in order to make direct use of the
continuum  $\chi$PT for the sake of the extrapolations. The domain
wall  fermion (DWF) formulation is well-suited in this
regard,  since it preserves exact flavor symmetry, and chiral symmetry
is only mildly broken. Its chiral symmetry breaking effect can be
quantitatively described by a small additive mass shift called the
residual mass, $\mres$. Recent work has shown~\cite{Lin:2006cf,
  Sharpe:2007yd}  that, to do chiral extrapolations for domain wall
fermions,   the only modification to the continuum $\chi$PT   is to
replace the  input quark mass by the sum of the input quark mass and
$\mres$,  leaving the number of low energy constants
unchanged, at least up to terms of ${\cal O}(ma)$ which can be viewed as next-to-next-to-leading order (NNLO). This is  in contrast to the cases of Wilson fermions or
staggered  fermions, where, at next-to-leading order, a few new low-energy constants need to be
introduced  to account for the chiral symmetry (Wilson) or flavor
symmetry  (staggered) breaking effects. 

One of the challenges of chiral extrapolations is that it still
remains  inconclusive what the radius of convergence is for $\chi$PT.
Previous results of domain wall fermion simulations~\cite{Lin:2006cf}
have evidence that $\chi$PT at  next-to-leadig order (NLO) is not sufficient to
describe pion masses heavier than 400 MeV. One question to ask is, how
light  should the pion masses (or quark masses) be for $\chi$PT to
achieve  the desired accuracy at NLO?  In this work we
present results for the pseudoscalar meson masses and decay constants
from recent domain  wall fermion simulations with 2+1 dynamical
flavors on the  $24^3\times64$ lattices at a fixed lattice spacing of
about  0.1 fm. The partially quenched pion masses in these simulations
are as light as 250 MeV, which gives us an opportunity to check if
$\chi$PT  is consistent with the lattice data at this lighter mass
range. The  agreement between the lattice data and the predictions of
$\chi$PT  in turn enables us to determine physical observables and the
light  quark masses with better controlled extrapolation errors than
na\"ive  linear fits.  In this proceedings we combine two talks given at
the  Lattice 2007 conference, and show our attempts to locate the mass
range  where $\chi$PT (SU(3)$\times$SU(3) and SU(2)$\times$SU(2)) is
applicable,  followed by the determinations of
$f_\pi$,  $f_K$ and the physical light (up/down and strange) quark masses. 
For other physical results obtained from these configurations see \cite{Boyle:Proc} and references therein.

\section{Numerical Details}
The gauge configurations on the $24^3\times64$ lattices were generated
using the same parameters as the previous simulations on the
$16^3\times32$ lattices~\cite{Allton:2007hx}. Specifically, we used
the Iwasaki gauge action with $\beta = 2.13$. The extent of the fifth
dimension was $L_s = 16$, and the domain wall height was fixed to $aM_5
= 1.8$.  The dynamical strange quark mass, $a m_s = 0.04$,  was tuned
to be approximately its physical value, and four values of the light
dynamical quark mass, $a m_l$, were used to allow for the
extrapolations in the light quark mass limit. The rational hybrid
Monte Carlo (RHMC)  algorithm was applied to generate all the
ensembles. The details of the implementation for the RHMC were reported in ~\cite{Mawhinney:lat06}. The number of thermalized trajectories, in molecular dynamics time units, for the $am_l = 0.005, 0.01, 0.02$ and $0.03$ ensembles is 3600, 3600, 1760 and 1760, respectively. 


In order to make full use of the partially quenched $\chi$PT
 formulae~\cite{Sharpe:2000bc}, we calculated
hadron correlators with the input valence quark masses $am_{x,y} \in
 \{0.001, 0.005, 0.01, 0.02, 0.03, 0.04 \}$. The lightest input quark mass turns out to be about 1/10 of the strange quark mass when the residual mass is properly included. Focus will be given
to the two ensembles with lightest sea quark masses, $a m_l= 0.005$
 and 0.01, as these smaller quark masses are more likely to be within
 the regime where NLO $\chi$PT has reasonable convergence. For these two ensembles, all the non-degenerate meson correlators were constructed
from all the different combinations of the six valence quark masses using a Coulomb gauge fixed wall source (W) and either a wall or local sink (L) as part of 
our weak matrix element project~\cite{Antonio:2007pb}. The quark
 propagators used in these measurements were obtained from the sum of 
quark propagators computed from  periodic and anti-periodic boundary
 conditions, to eliminate the boundary effects from the
 backward-propagating states.  For clarity we will denote these
 correlators as ``W-P+A''. Additionally, degenerate hadron
 correlators with a Coulomb gauge fixed $16^3$ box 
source and a $16^3$-box  or local sink were also calculated on these ensembles. 
Note that correlators constructed from a box source and a box sink violate translational invariance, therefore 
zero-momentum projection can not be guaranteed. 
We thus summed over the correlators with 
all the possible choices for the box sink to achieve the zero-momentum projection. 
The correlators with a box source
 were
 found to have better overlap with the ground states of the baryons,
 and were used to extract the mass of the $\Omega^-$ baryon, which we will
 utilize  to set the lattice scale for our simulations. The
 measurements were done on 90  gauge configurations on the $am_l =
 0.005$ and 0.01 ensembles, and about 45 on the $am_l =
 0.02$ and 0.03 ensembles, with two adjacent measurements separated by
 40 molecular dynamics time units. 
For each type of measurement we used two different source locations to reduce the fluctuations within the gauge configurations.  The small number of measurements on each ensemble does not allow us to 
study the autocorrelation time reliably. However, we have checked that blocking the data in intervals of 40 or 80 molecular dynamics time units does not
change the statistical errors significantly,  which is consistent with the study on the smaller volume~\cite{Allton:2007hx}. 
Thus in the following analysis, we choose to block the data into intervals of 80 molecular dynamics time units for the $a m_l = 0.005$ and 0.01 ensembles 
where 90 measurements are available, and into intervals of 40 molecular dynamics time units for the 0.02 and 0.03 ensembles, leaving approximately 45 jackknife samples for each of the four ensembles.

\section{Data Analysis}
\subsection{The Residual Mass $\mres$ and Axial Current Renormalization $Z_A$}
\begin{figure}[htbp]
  \hfill
  \begin{minipage}[t]{.45\textwidth}
    \begin{center}  
      \includegraphics[width=\textwidth]{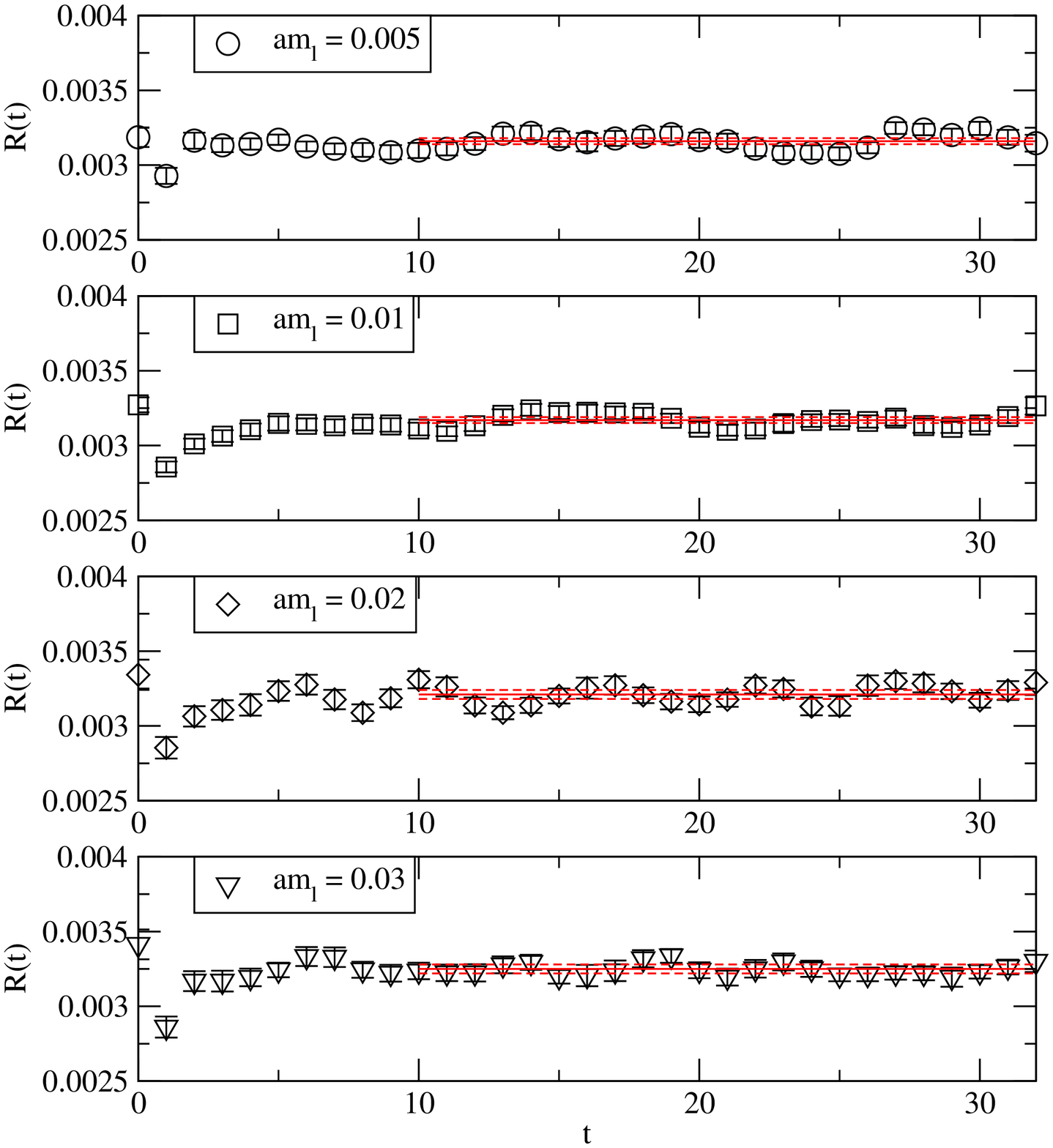}
      \vspace*{\closercaption}
      \caption{The ratio $R(t)$ used in the determination of the residual mass at the unitary points.}
      \label{fig:mres}
    \end{center}
  \end{minipage}
  \hfill
  \begin{minipage}[t]{.45\textwidth}
    \begin{center}  
      \includegraphics[width=\textwidth]{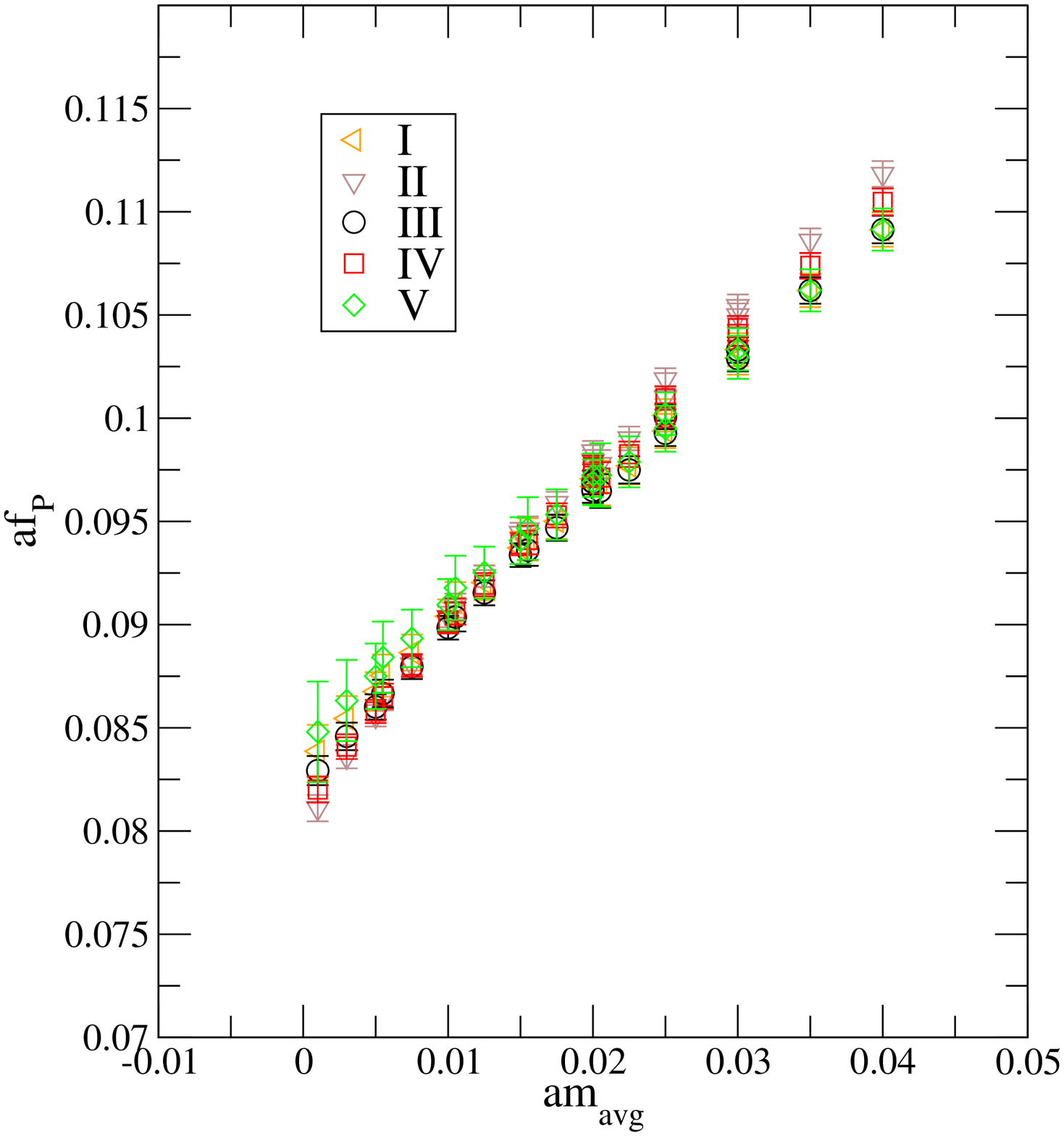}
      \vspace*{\closercaption}
      \caption{Results for $af_P$ using different methods on the $am_l = 0.005$ ensemble.}
      \label{fig:fpi}
    \end{center}
  \end{minipage}
  \hfill
\end{figure}

As the gauge coupling of this large volume simulation is identical to the $16^3\times 32$ simulations in \cite{Allton:2007hx}, we expect the residual mass $\mres$ and the axial current renormalization $Z_A$ 
to be consistent with the results therein up to possible finite volume
effects.  The residual mass is determined from the ratio~\cite{Antonio:2006px}
\begin{equation}
   R(t) = 
	  \frac{\langle \sum_{\vec{x}} J^a_{5q} (\vec{x}, t)
	    \pi^a(\vec{0},0) \rangle } 
	       {\langle \sum_{\vec{x}} J^a_5(\vec{x}, t) 
		 \pi^a(\vec{0},0) \rangle }, 
	       \label{eq:mres_ratio}
\end{equation}
where $J_{5q}$ is a point-split operator for domain wall fermions. Figure~\ref{fig:mres} shows the results of $R(t)$ at the four unitary points with $am_x = am_y = am_l$. The horizontal lines represent the fit to a constant from $t = 10$ to 32 for each quark mass, determining $a\mres'(am_l)$. The mass-independent residual mass is given by evaluating $a\mres'(am_l)$ at $am_l = 0$, and we have 
\begin{equation}
 a\mres = 0.00315(2). 
\end{equation}

The axial current renormalization constant $Z_A$ relates the local axial vector current \linebreak[9] $ \bar{q}(x)\tau^a \gamma_4 \gamma_5 q(x)$ to the conserved axial vector current of domain wall fermions~\cite{Blum:2000kn}. It is determined from the ${\cal O}(a^2)$ improved ratio as described in Ref.~\cite{Blum:2000kn}. Similar to the residual mass, we compute the value of $Z_A$ at each unitary quark mass, and extrapolate to the chiral limit at $a m_l = - a \mres$, obtaining 
\begin{equation}
 Z_A = 0.7161(1).
\end{equation}
The results for $\mres$ and $Z_A$ are consistent with the previous results in the small volume, indicating no measurable finite volume effects are present for these quantities.

\subsection{Pseudoscalar Meson Masses and Decay Constants}
This section is devoted to the details of our fitting procedures to obtain the light pseudoscalar meson masses, $m_P$. There are two types of interpolating operators which overlap with the pseudoscalar meson state: $P(x,y) \equiv \bar{q}(x) \tau^a \gamma_5 q(y)$ and $A(x,y) \equiv \bar{q}(x)\tau^a \gamma_4 \gamma_5 q(y)$, where the quark fields may 
have different smearings. To minimize systematic errors arising from
different characteristics of the operators, we chose to fit all the
available W-P+A correlators simultaneously to obtain a common mass and
an amplitude for each correlator.  Since the correlators are measured
on the same gauge background, in principle we should take into account
correlations among different correlators and  different time slices of
the same correlator. However, the fit range for a typical simultaneous fit
is as large as 220 time slices.  Having only 45 jackknife samples is not  enough to resolve the  covariance matrix. Thus the correlated fits cannot be carried out, and we restrict ourselves to uncorrelated fits.  One caveat of the uncorrelated fits is that  $\chi^2$/d.o.f.\ from the uncorrelated fits does not follow the correct $\chi^2$ distribution, and do not reliably imply the goodness of the fits. 

The following five correlators were included in the simultaneous fits:  
\begin{equation} 
   \langle A^L (t) A^W(0) \rangle ,     \langle P^L (t) P^W(0) \rangle ,   \langle A^L (t) P^W(0) \rangle  ,  \langle P^W (t) P^W(0) \rangle , \, \, {\rm and}\,\,     \langle A^W (t) P^W(0) \rangle , 
\end{equation}
where the superscripts indicate the smearing of the source or
sink, with $W$ being the Coulomb gauge fixed wall and $L$ being the local operator. Each simultaneous fit gives a common mass $am_P$, and one
amplitude for each correlator, labelled as ${\cal A}_{AA}^{LW}, {\cal
  A}_{PP}^{LW}, {\cal A}_{AP}^{LW}, {\cal A}_{PP}^{WW}$ and ${\cal
  A}_{AP}^{WW}$, respectively. There are five different ways to
determine the pseudoscalar meson decay constant\footnote{Our definition for the decay constant is such that the physical value of $f_\pi$ is about 130 MeV.}, $af_P$, using these
amplitudes: 
\begin{eqnarray}
 (af_P)^2 & = & \frac{2 Z_A^2}{(am_P) V} \frac{{\cal A}_{AA}^{LW} \,
 {\cal A}_{AP}^{LW}}{{\cal A}_{AP}^{WW}}\, {\rm(I)},\,
 \frac{8[\frac{1}{2}(am_{x}+am_{y}) + a\mres]^2}{(am_P)^3 V} \frac{{{\cal
 A}_{PP}^{LW}}^2}{{\cal A}_{PP}^{WW}}\,{\rm (II)}, \,
 \frac{2 Z_A^2}{(am_P) V} \frac{{{\cal A}_{AP}^{LW}}^2}{{\cal A}_{PP}^{WW} } \,
{\rm (III)},\nonumber
\\
 & &4 Z_A \frac{\frac{1}{2}(am_x+am_y) + a\mres}{(am_P)^2 V} \frac{{\cal
     A}_{PP}^{LW} {\cal A}_{AP}^{LW}}{{\cal A}_{PP}^{WW}}\,\ {\rm (IV)},\,\,\,
     {\rm and} \,\,\,
     \frac{2 Z_A^2}{(am_P) V}\frac{{{\cal A}_{AA}^{LW}}^2\, {\cal
     A}_{PP}^{WW}}{{{\cal A}_{AP}^{WW}}^2}\ \,{\rm (V)},
\label{eq:fpi}
\end{eqnarray}
where $V\equiv (aL)^3$ is the spatial volume of the lattice. These ratios are calculated under a standard jackknife procedure 
to take into account correlations among different amplitudes. 
Note that not  all of these methods are independent, but some of them
may produce statistically more accurate results than the others due to
different characteristics of the correlators.  (II)  and (IV) in fact come from 
the translation from pseudoscalar density to axial vector current
using the axial Ward identity~\cite{Blum:2000kn},  hence the residual
mass $\mres$ is required.  The results for $af_P$ of the $am_l =
0.005$ ensemble from all of these different methods are shown in
Fig.~\ref{fig:fpi}. As we can see, they all give consistent results
except that methods (II) and (IV) give slightly higher results than
the rest at large masses, which may indicate different scaling errors resulting from
the use of different correlators. In the following analysis, we use results from (III) since it gives the smallest statistical error.

\section{Chiral fits: SU(3)$\times$SU(3) and SU(2)$\times$SU(2)}

In this section we will discuss our attempts to fit the obtained meson masses and decay constants to formulae predicted by partially quenched chiral perturbation theory (PQ$\chi$PT). (For similar fits for the kaon bag parameter $B_K$ measured on the same lattice configurations see \cite{Antonio:2007pb,AntonioCohen:Proc}.) Using PQ$\chi$PT for three quark masses, corresponding to unquenched $\SU(3)\times\SU(3)$ $\chi$PT, up to NLO it will turn out that the data at our higher quark masses is not well described by the applicable formulae (Sect.~\ref{subsec:su3Fits}). Therefore, in Sect.~\ref{subsec:su2Fits} we will perform NLO $\SU(2)\times\SU(2)$ fits, dropping terms of order $(m_l/m_s)^2$.

\subsection{\label{subsec:su3Fits}SU(3)$\times$SU(3) Chiral Fits}
The most natural approach to fit our data from the $N_f=2+1$ ensembles is to use $\SU(3)\times\SU(3)$ $\chi$PT or its partially quenched variant, describing the dependence of the meson masses and decay constants on the two (in our case degenerate) light quark masses and the heavier strange quark mass by introducing chiral fit parameters to leading order (LO: $B_0$, $f_0$) and next-to-leading order (NLO: $L_{4,5,6,8}$), where the latter are commonly referred to as \textit{Gasser-Leutwyler parameters} or \textit{low energy constants} (LECs). From the general formulae given in \cite{Sharpe:2000bc} the $N_f=2+1$ case has been worked out, see, e.g., \cite{Lin:2006cf}.

\begin{figure}
\begin{center}
\begin{minipage}{.5\textwidth}
\begin{center}
\includegraphics[angle=-90, width=.9\textwidth]{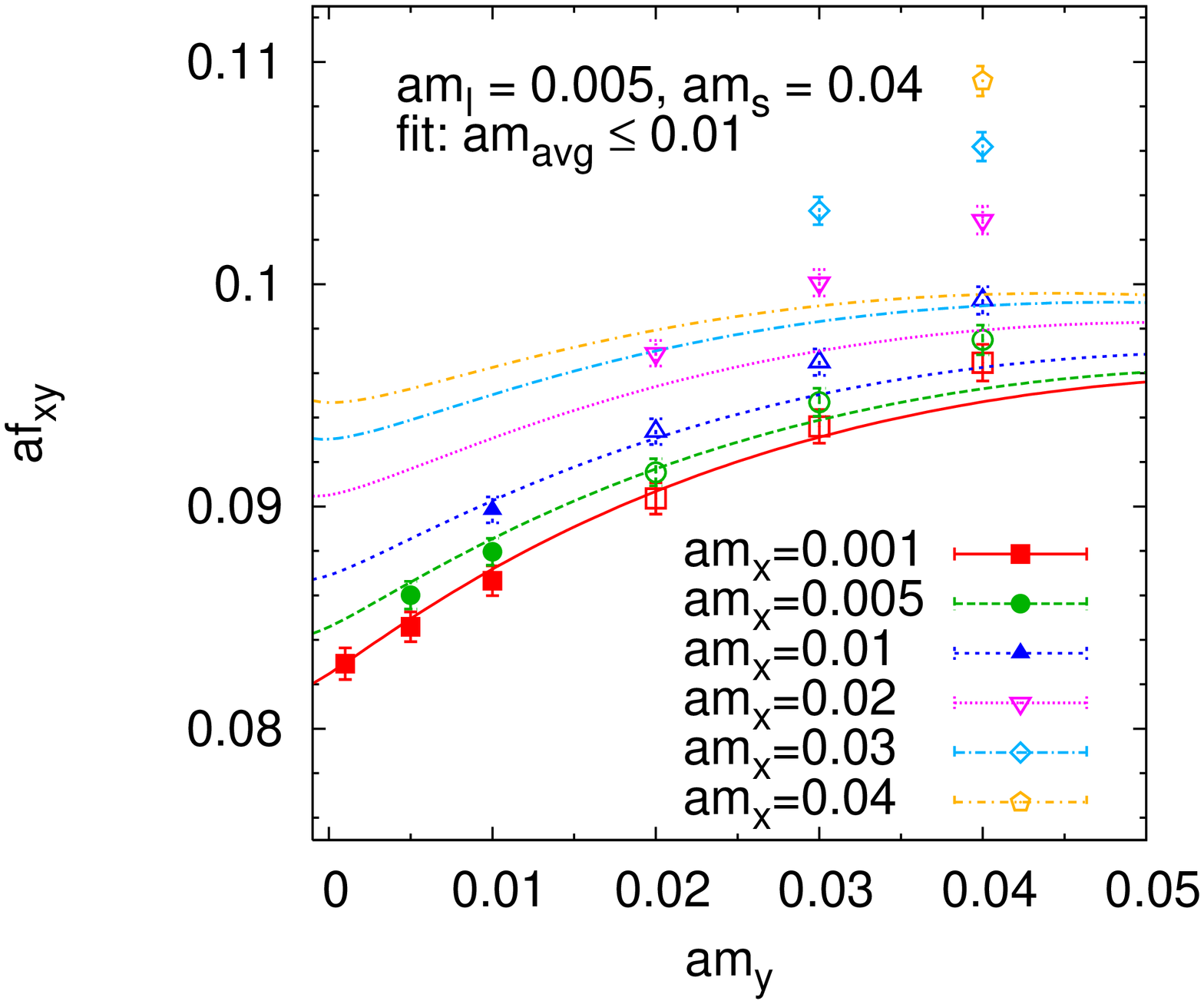}\\
\includegraphics[angle=-90, width=.9\textwidth]{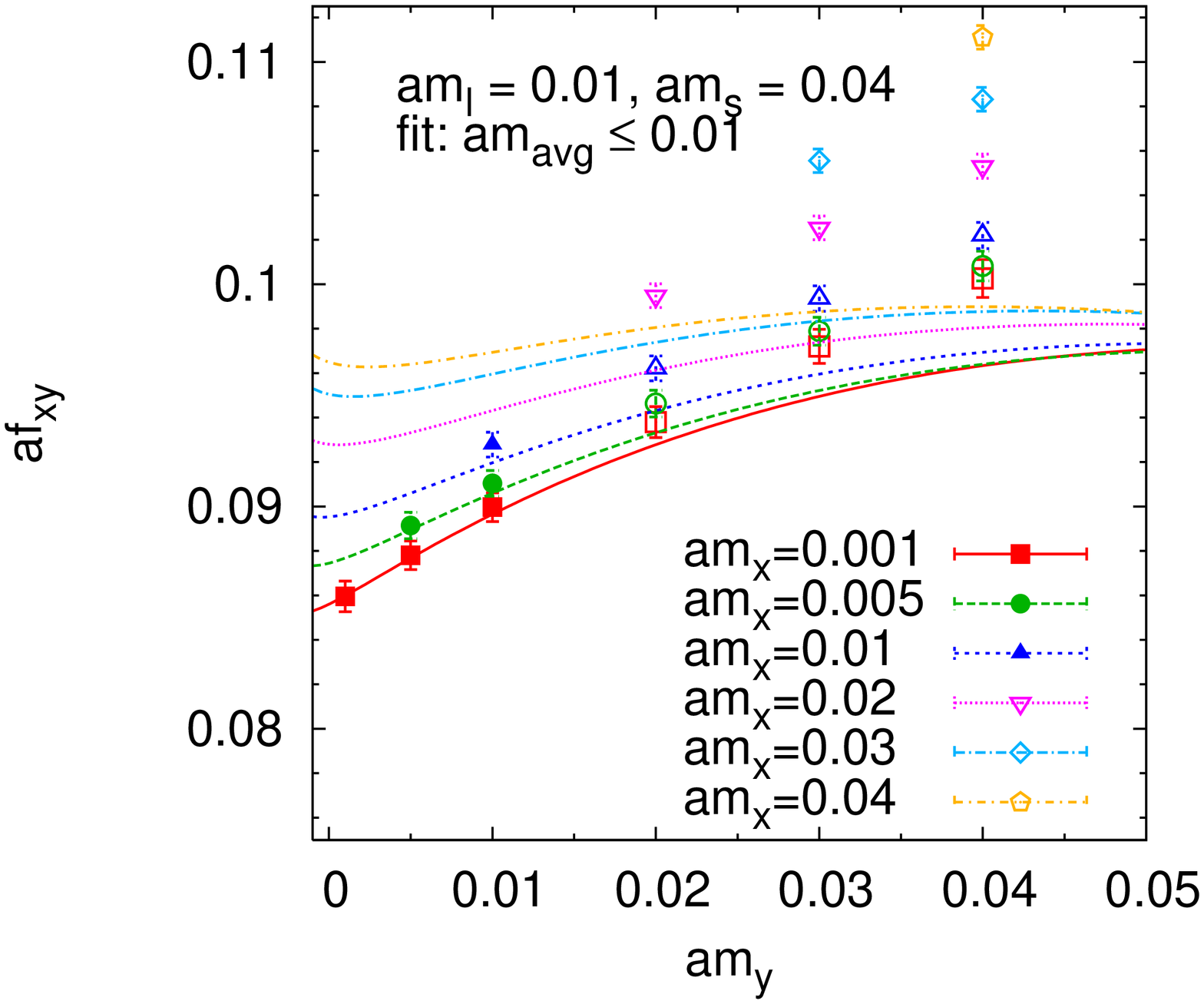}
\end{center}
\end{minipage}%
\begin{minipage}{.5\textwidth}
\begin{center}
\includegraphics[angle=-90, width=.87\textwidth]{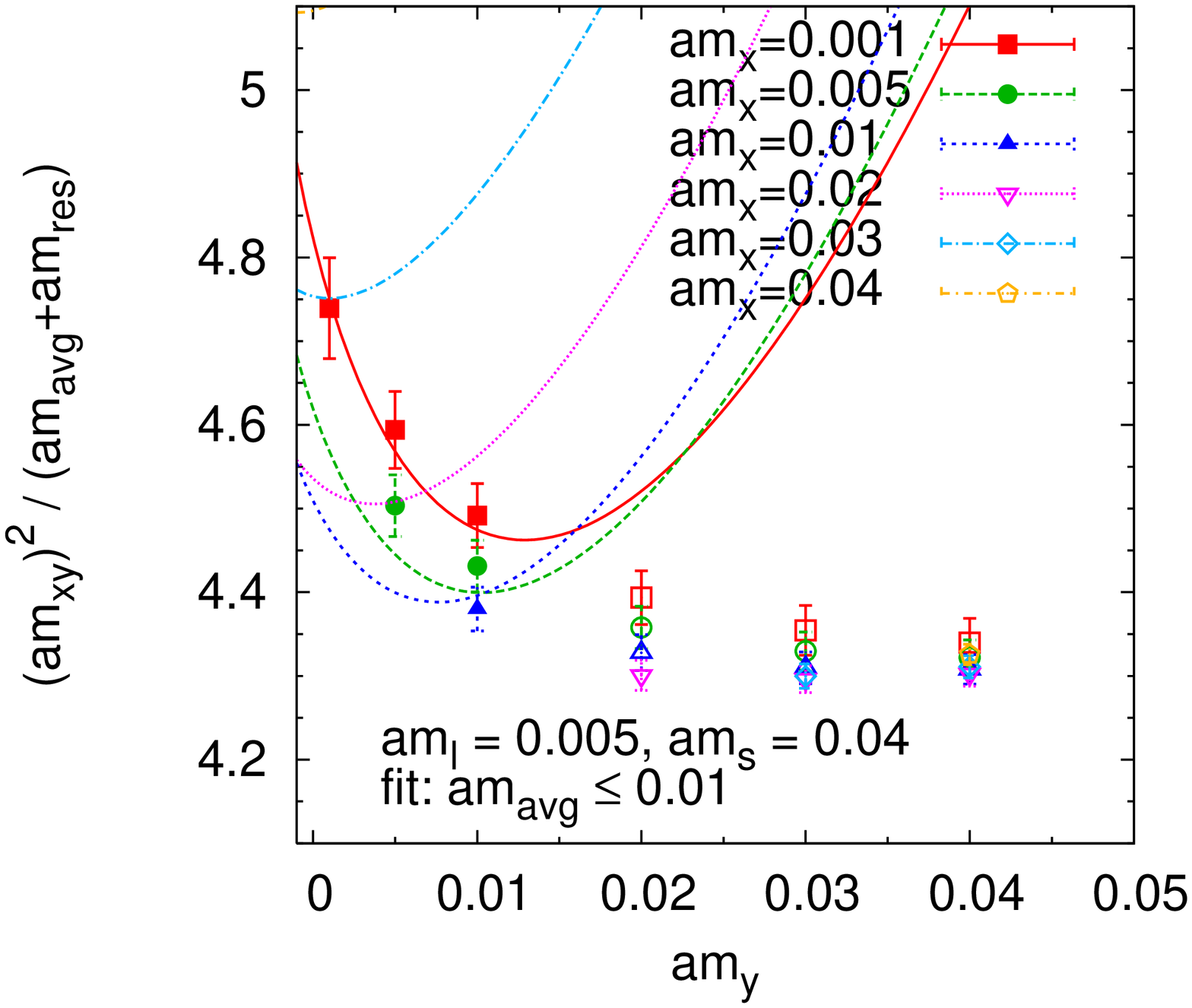}\\
\includegraphics[angle=-90, width=.87\textwidth]{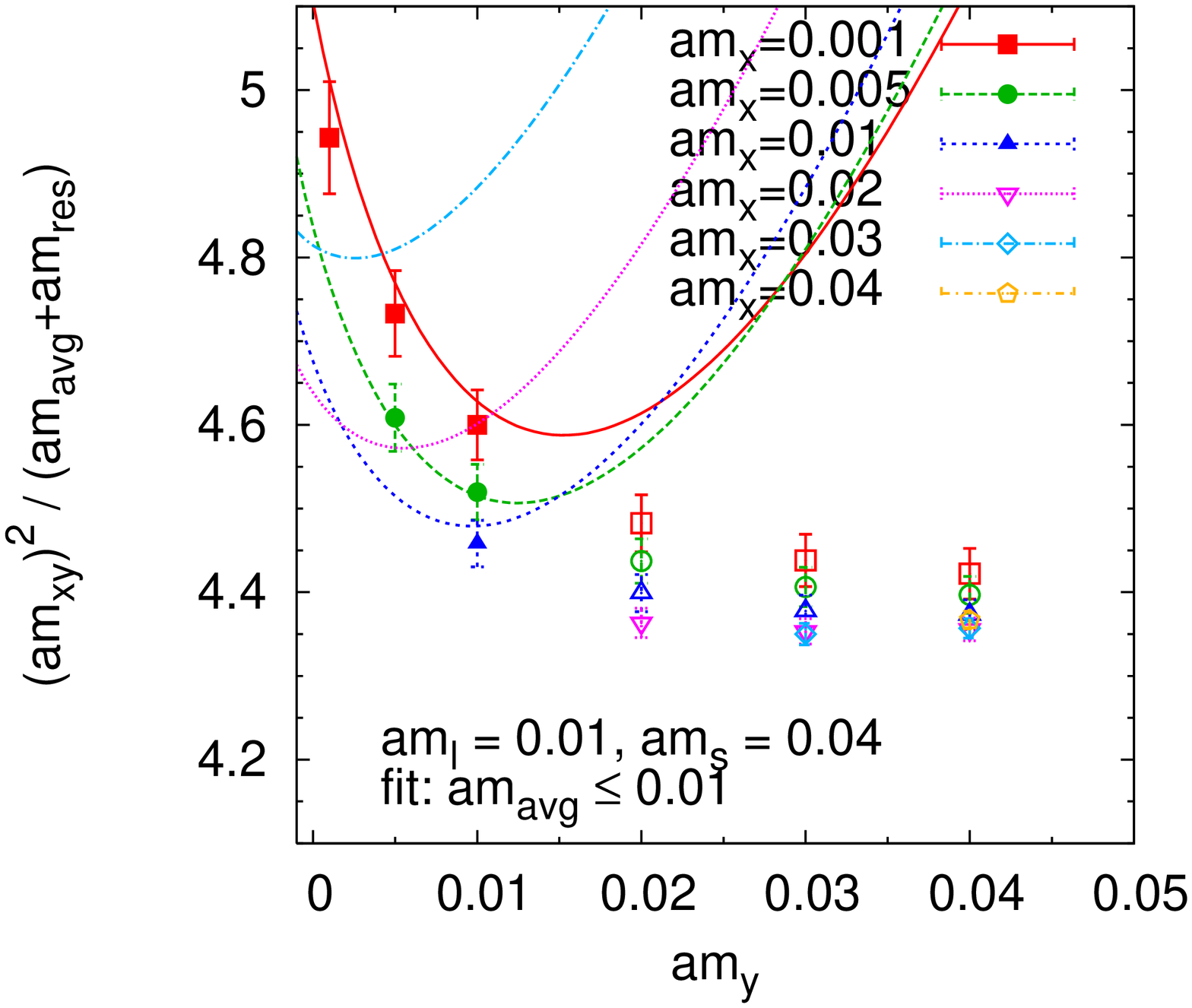}
\end{center}
\end{minipage}
\end{center}
\vspace*{\closercaption}
\caption{\label{fig:SU3fits}Combined $\SU(3)\times \SU(3)$ fits for the meson decay constants \textit{(left panels)} and masses \textit{(right panels)} at two different values for the light sea quark mass, valence mass cut $am_{\rm avg}\leq0.01$. Points marked by \textit{filled symbols} were included in the fit, while those with \textit{open symbols} were excluded.}
\end{figure}

\begin{figure}
\begin{center}
\begin{minipage}{.5\textwidth}
\begin{center}
\includegraphics[angle=-90, width=.9\textwidth]{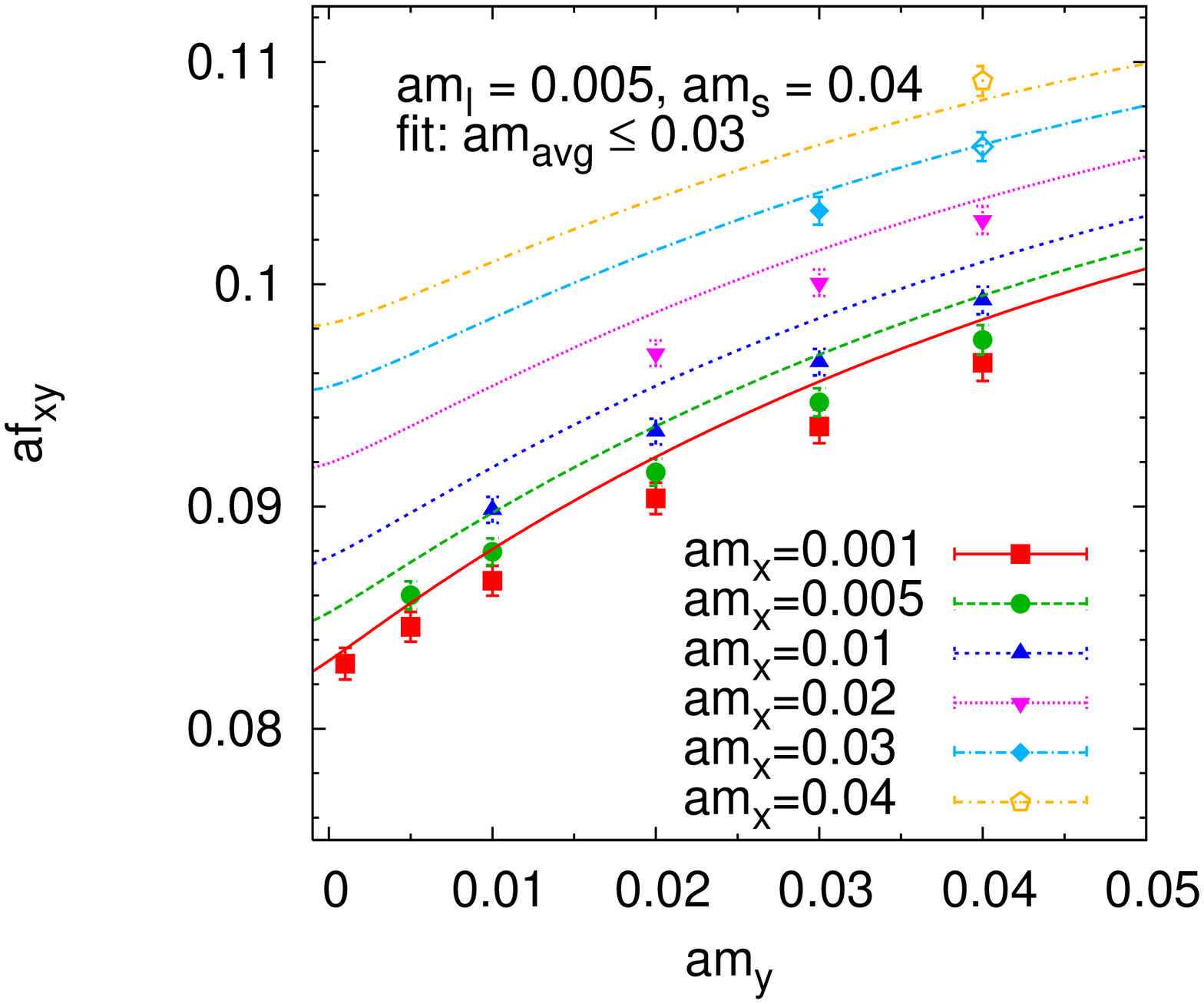}\\
\includegraphics[angle=-90, width=.9\textwidth]{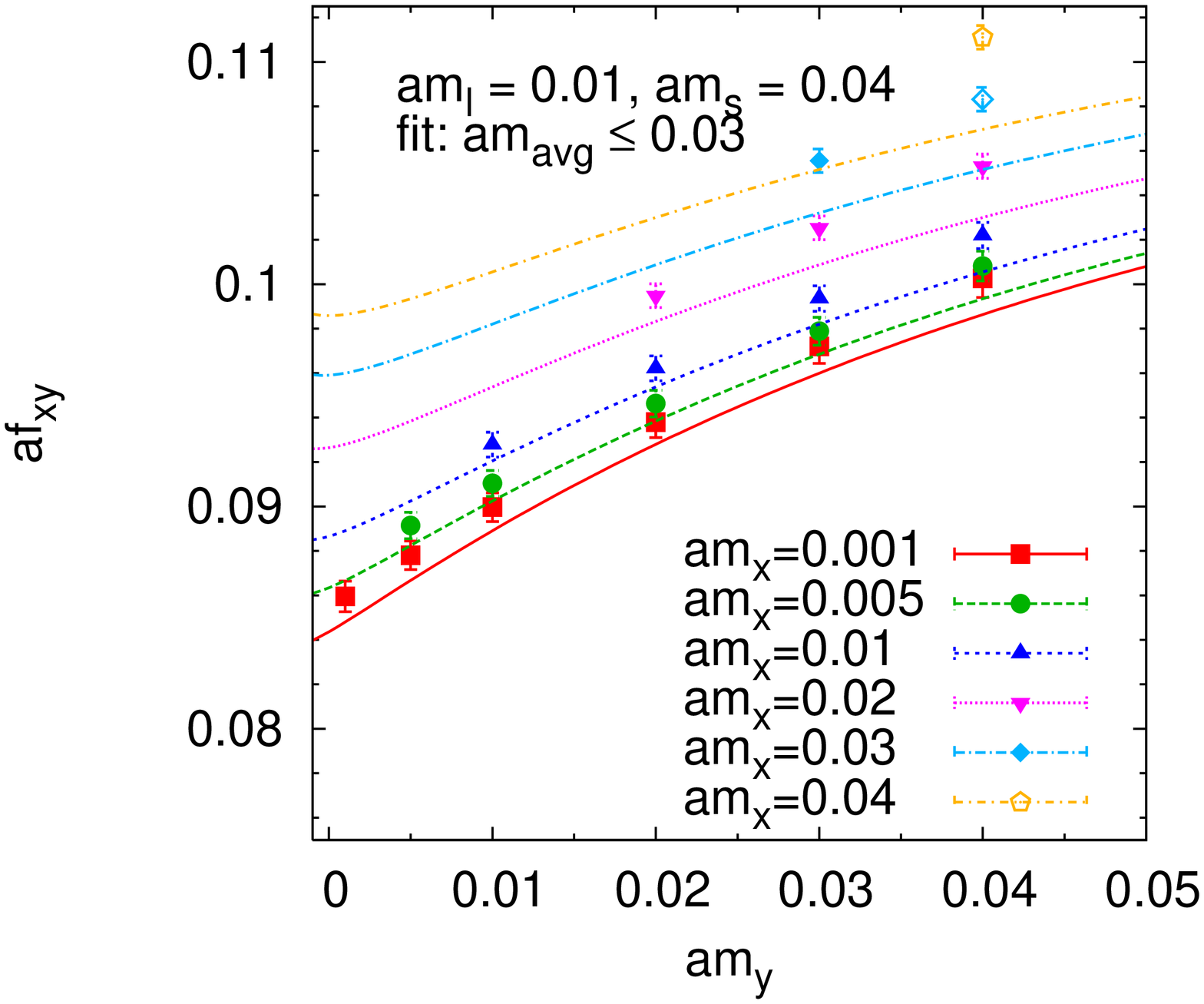}
\end{center}
\end{minipage}%
\begin{minipage}{.5\textwidth}
\begin{center}
\includegraphics[angle=-90, width=.87\textwidth]{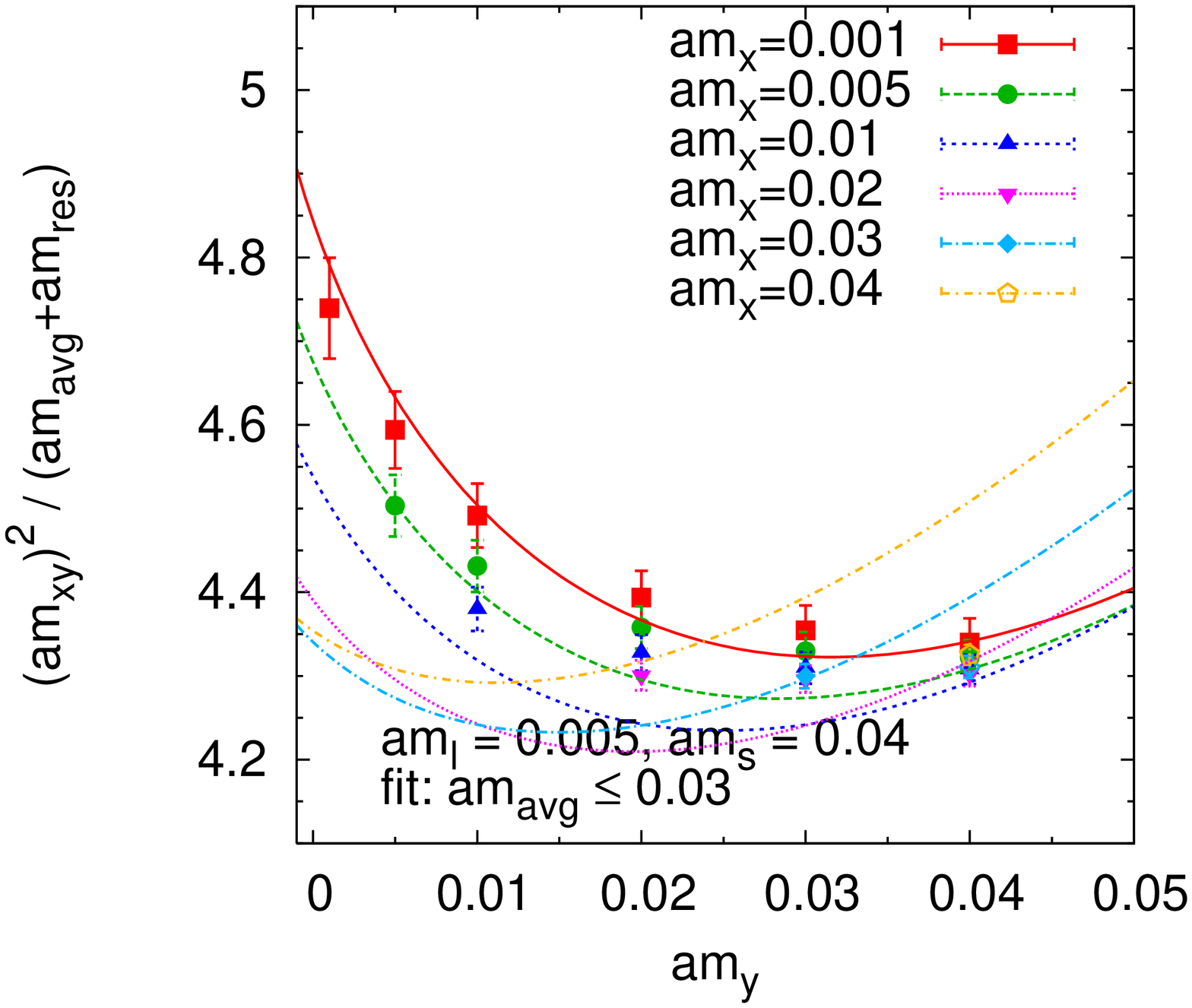}\\
\includegraphics[angle=-90, width=.87\textwidth]{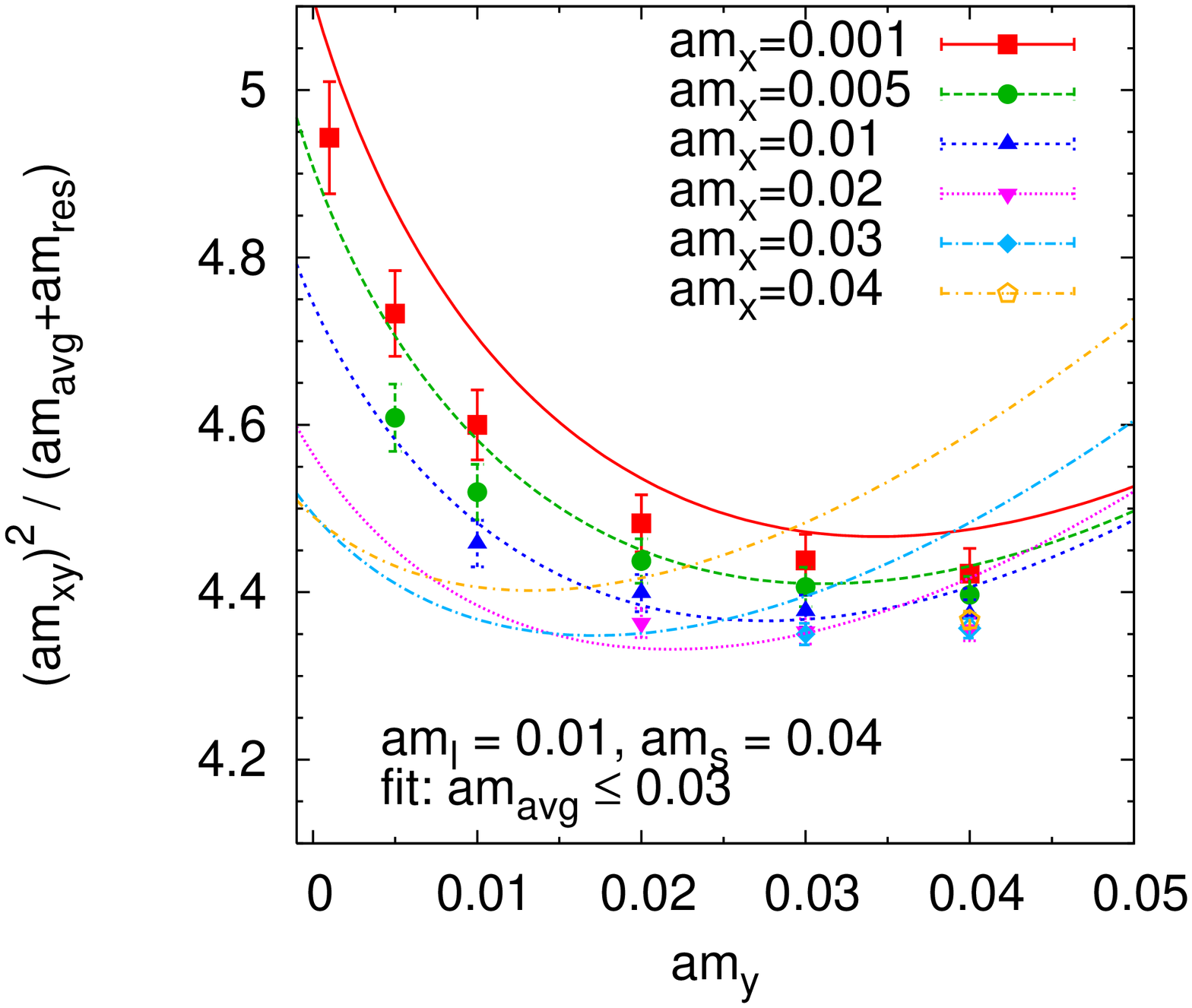}
\end{center}
\end{minipage}
\end{center}
\vspace*{\closercaption}
\caption{\label{fig:SU3fits_bad}Combined $\SU(3)\times \SU(3)$ fits for the meson decay constants \textit{(left panels)} and masses \textit{(right panels)} at two different values for the light sea quark mass, valence mass cut $am_{\rm avg}\leq0.03$. Points marked by \textit{filled symbols} were included in the fit, while those with \textit{open symbols} were excluded.}
\vspace*{\afterFigure}
\end{figure}

When appying these fit forms to our data, we found that using (PQ)$\chi$PT to NLO does not describe our data well up to meson masses comparable to the kaon mass or---equivalently---up to an average quark mass of half the strange quark mass. We performed combined fits to $af_{xy}$ and $(am_{xy})^2$, meaning the decay constant or mass squared of a meson built from valence quarks with masses $m_x$ and $m_y$, using the two ensembles with dynamical light quark masses of $am_l=0.005$ and $0.01$. A reasonable $\chi^2/{\rm d.o.f.}$ could only be obtained by imposing a cut in the average valence quark mass of $am_{\rm avg}\equiv(am_x+am_y)/2 \leq 0.01$; fits with such a cut are shown in Fig.~\ref{fig:SU3fits}, while the fit parameters are given in Tab.~\ref{tab:fitres}. There we conveniently quote the scale-dependent LECs at two commonly used chiral scales of $\Lambda_\chi=1.0\,{\rm GeV}$ and 770 MeV. Also included in the table are phenomenological estimates for the LECs from \cite{Bijnens:2007yd} and references therein. Our results show agreement with their NNLO fit values. 
In Fig.~\ref{fig:SU3fits_bad} we show fits with the cut chosen to be $am_{\rm avg}\leq 0.03$. The fits miss almost all the data points inside the fitting range. Therefore, we conclude that NLO-$\chi$PT fits are not reliably applicable in a mass range up to the kaon mass. If one were to extract just the pion sector quantities, i.e., just the physical $f_\pi$, $m_\pi$, and the physical light quark mass, from the fit results with the low mass cut, one still would include the terms proportional to the strange quark mass. Ideally, one would like to use $\chi$PT to guide the interpolation to the physical value of the latter. However, since we saw that at such a quark mass the fits deviate substantially from the data, this procedure has to be seen as an unsafe or at least questionable one. For the same reasons, a meaningful extraction of quantities in the kaon sector is impossible within this approach.

One could try to extend the range of validity of $\chi$PT by going
from NLO to NNLO. The complete formulae are available in the
literature \cite{Bijnens:2006jv}. However, this would introduce 
much more LECs than the number of independent data points which are currently available to us. 
In addition, under these circumstances, we would not be able to establish whether such a NNLO fit was itself appropriate for this mass range. Were this kinematic region outside the domain of validity of $\chi$PT, such NNLO terms may not correctly describe our results.
Instead we followed a different ansatz, namely to base the fit formulae just on the (approximate) chiral symmetry within the light quark doublet, as will be described in the next subsection.

\begin{table}
%
\newsavebox{\refNPR}
\savebox{\refNPR}{\ref{subsec:npr}}
\caption{\label{tab:fitres}Fitted parameters from different fits with a valence mass cut $am_{\rm avg}\leq0.01$. For each fit the LECs are quoted at two different scales $\Lambda_\chi$. (Note: the value of $B_0$ depends on the renormalization scheme like the quark masses: to obtain $B_0$, e.g., in the $\overline{\rm MS}(2\,{\rm GeV})$ scheme, one has to divide the here quoted values by $Z_m^{\overline{\rm MS}}(2\,{\rm GeV})$ from Sect.~\usebox\refNPR.) Also included are the phenomenological estimates from \cite{Bijnens:2007yd}. Errors on $L_8$ and $L_5$ in \cite{Bijnens:2007yd}  are added by quadrature to give the error on $2L_8-L_5$. } 
\begin{center}
\begin{tabular}{rR*{4}{C}}\hline
&\multicolumn{1}{C}{\Lambda_\chi} & (2L_8-L_5) & L_5 & (2L_6-L_4) & L_4 \\[3pt]\hline
\multicolumn{6}{L}{\SU(3)\times\SU(3):\:aB_0 = 2.35(16),\, af_0 =  0.0541(40)}\\[2pt] 
                   & 1\,{\rm GeV}
                   & 5.19(45)\sci{-4} & 2.51(99)\sci{-4} & -4.7(4.2)\sci{-5} & -6.7(8.0)\sci{-5} \\
                   & 770\,{\rm MeV}  
                   & 2.43(45)\sci{-4} & 8.72(99)\sci{-4} & -0.1(4.2)\sci{-5} & 1.39(80)\sci{-4} \\[3pt]
\multicolumn{6}{l}{$\SU(3)\times\SU(3)$ LECs from \cite{Bijnens:2007yd}:}\\
 NLO  &  770\,{\rm MeV} & 5.4\sci{-4} & 14.6\sci{-4} & \equiv 0 & \equiv 0\\
 NNLO &  770\,{\rm MeV} & 2.3(3.8)\sci{-4} & 9.7(1.1)\sci{-4} & \equiv 0 & \equiv 0\\[5pt]

\multicolumn{6}{L}{\SU(2)\times\SU(2):\: aB_0 = 2.414(61),\, af_0 =  0.0665(21)}\\[2pt] 
                   & 1\,{\rm GeV}  
                   & 4.64(43)\sci{-4} & 5.16(73)\sci{-4} & -7.1(6.2)\sci{-5}&  1.3(1.3)\sci{-4}\\
                   & 770\,{\rm MeV}  
                   & 5.0(4.3)\sci{-5} & 9.30(73)\sci{-4} & 3.2(6.2)\sci{-5} & 3.3(1.3)\sci{-4} \\\hline
\end{tabular}
\end{center}
\end{table}

\subsection{\label{subsec:su2Fits}SU(2)$\times$SU(2) Chiral Fits}
First, we will purely focus on the pion sector. 
By applying NLO $\SU(2)\times\SU(2)$ (PQ)$\chi$PT, where terms of order
$(m_l/m_s)^2$ have been dropped, the strange quark mass will not explicitly enter the fit formulae. The dynamical strange quark mass present in our simulations acts as a background field and is therefore implicitly contained in the $\SU(2)\times\SU(2)$ LECs. 
Of course, in that way we will not be able to correct for the difference between the dynamical strange quark mass value, which was fixed during the generation of the gauge configuration ($am_s=0.04$), and its value at the physical point. As we shall see later on after extracting the physical $am_s^{\rm phys}$, this difference amounts to about 15 percent.

\begin{figure}
\begin{center}
\begin{minipage}{.5\textwidth}
\begin{center}
\includegraphics[angle=-90, width=.9\textwidth]{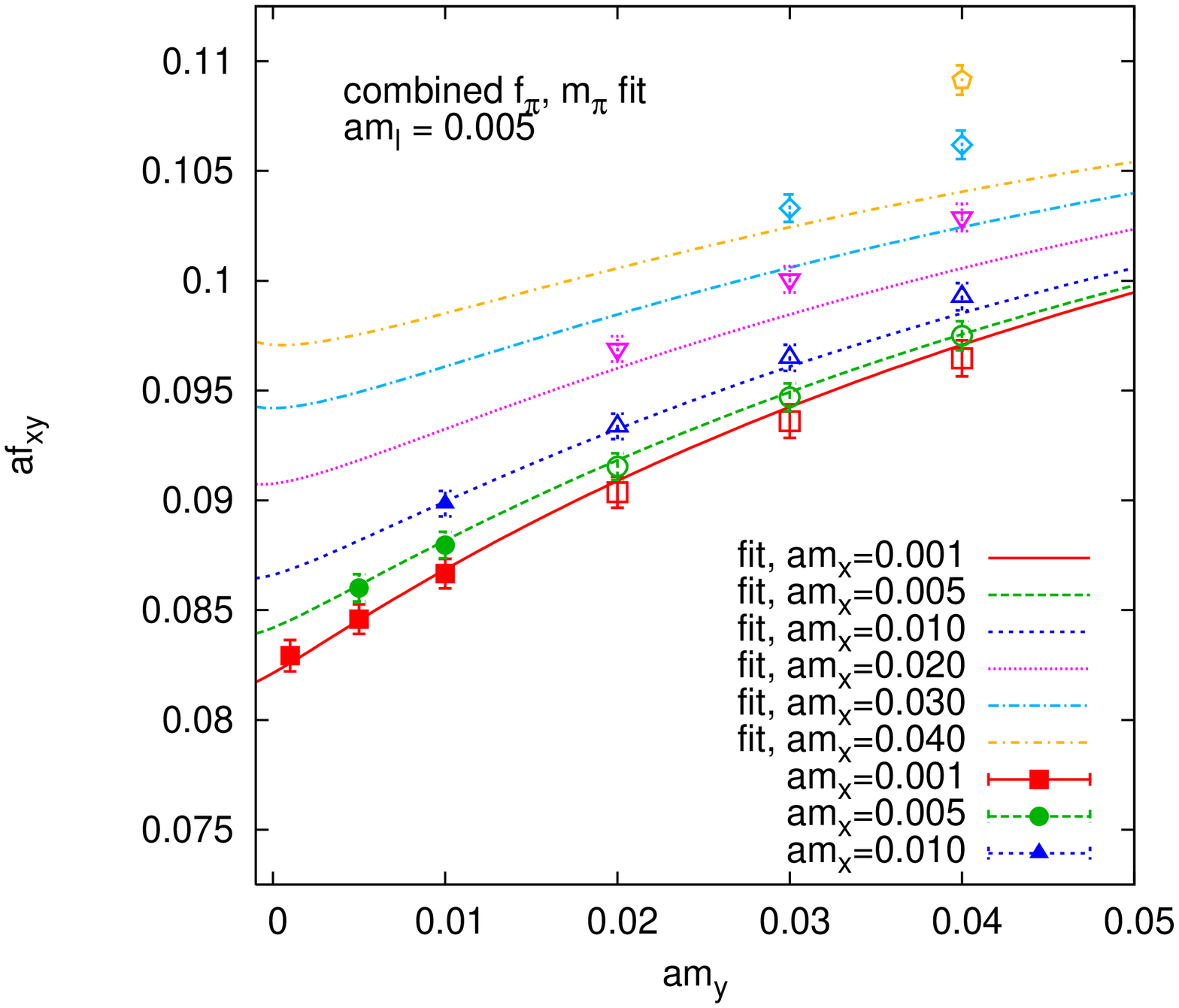}\\
\includegraphics[angle=-90, width=.9\textwidth]{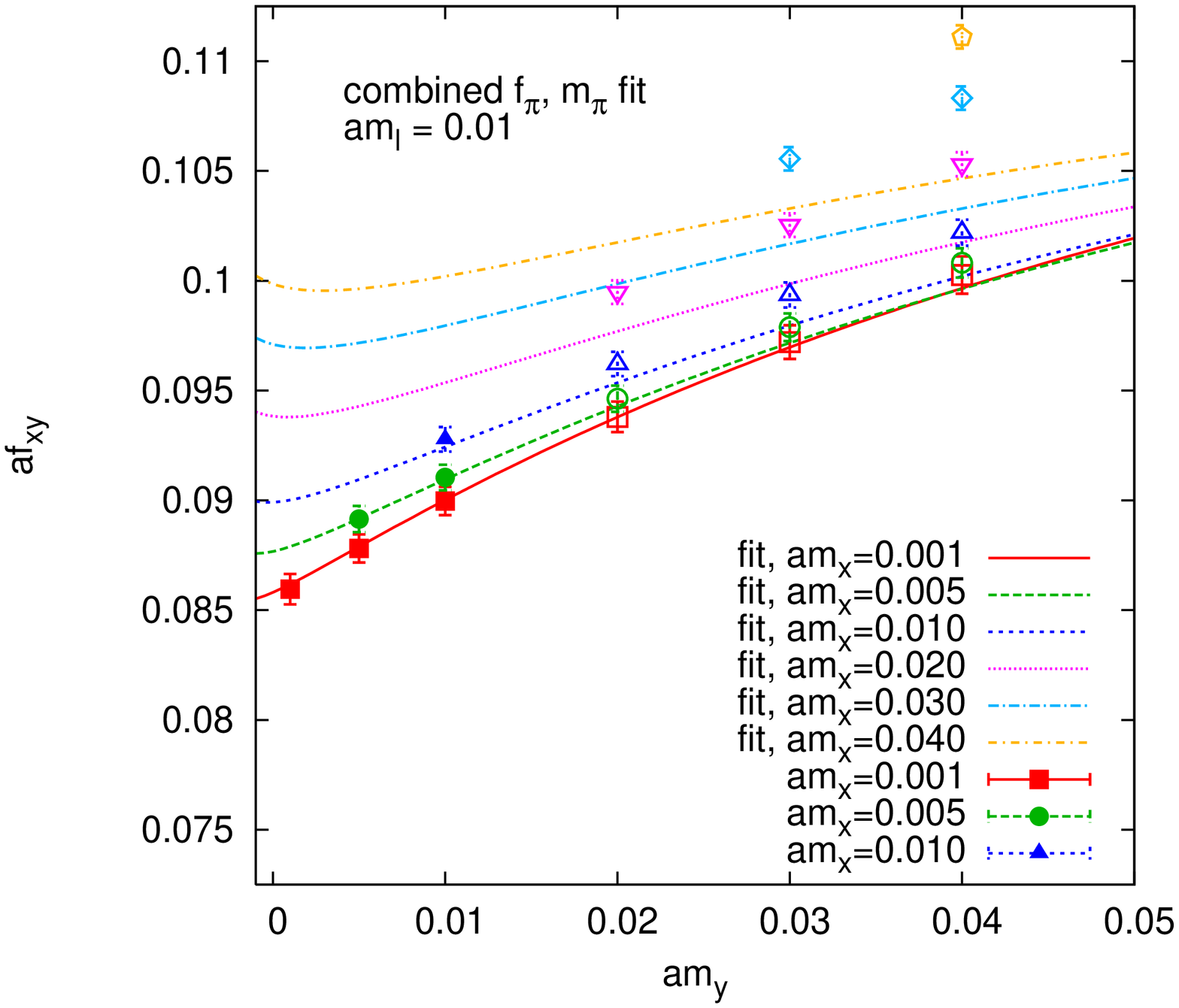}
\end{center}
\end{minipage}%
\begin{minipage}{.5\textwidth}
\begin{center}
\includegraphics[angle=-90, width=.87\textwidth]{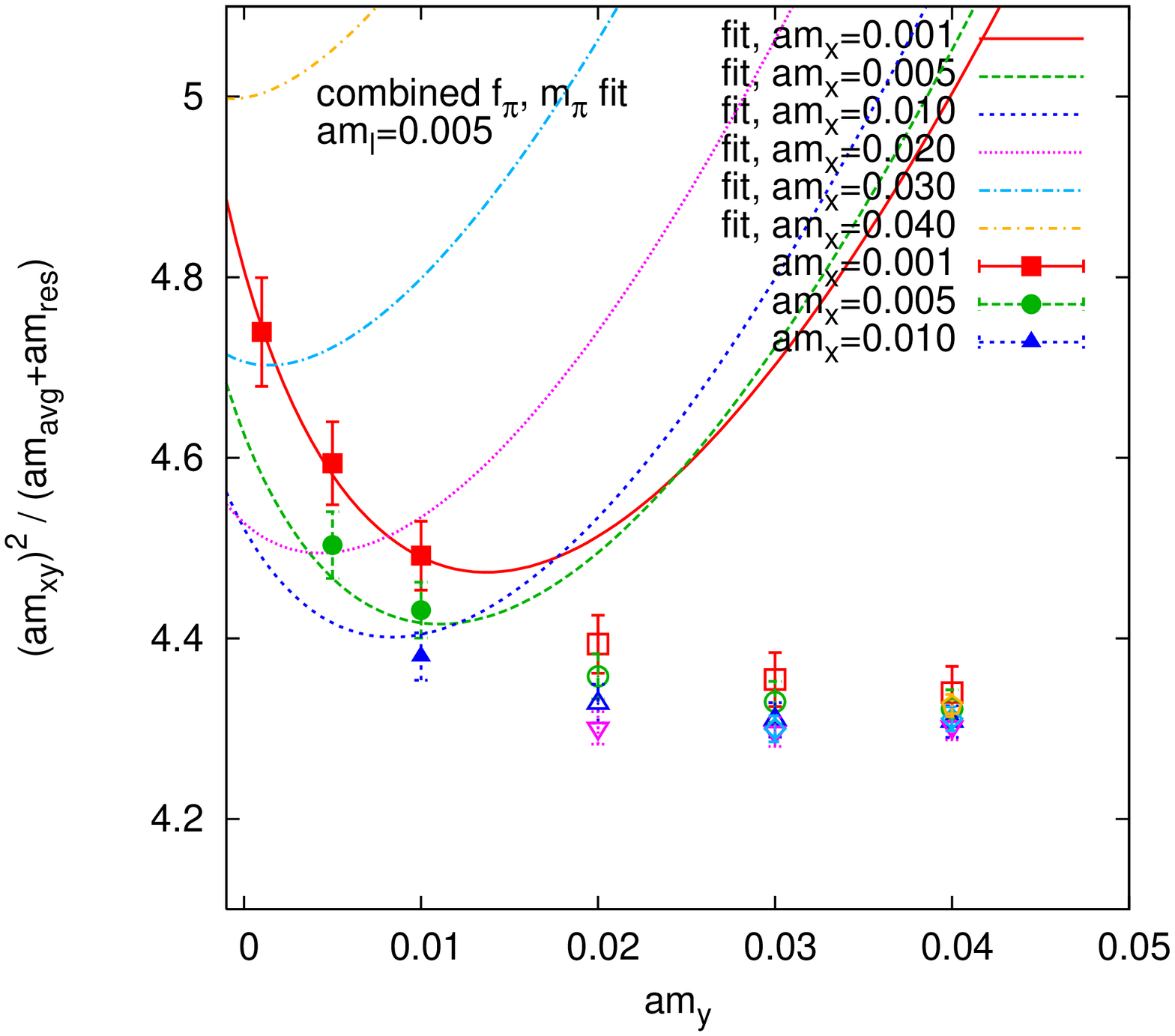}\\
\includegraphics[angle=-90, width=.87\textwidth]{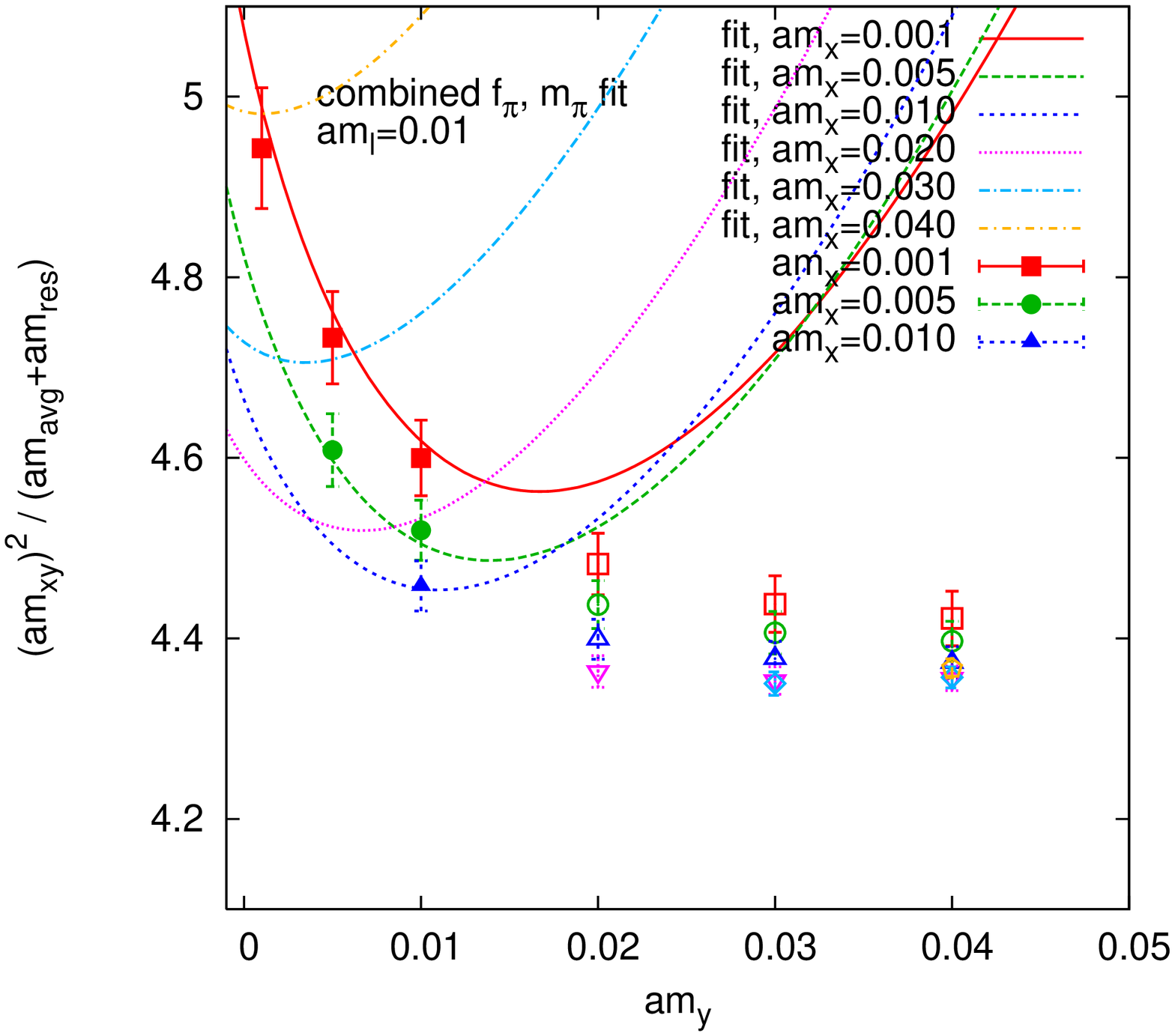}
\end{center}
\end{minipage}
\end{center}
\vspace*{\closercaption}
\caption{\label{fig:SU2fits}Combined $\SU(2)\times \SU(2)$ fits for the meson decay constants \textit{(left panels)} and masses \textit{(right panels)} at two different values for the light sea quark mass, valence mass cut $am_{\rm avg}\leq0.01$. Points marked by \textit{filled symbols} were included in the fit, while those with \textit{open symbols} were excluded.}
\vspace*{\afterFigure}
\end{figure}

The fit formulae for $N_f=2$ are derived starting from \cite{Sharpe:2000bc}, too, as has been done, for instance, in \cite{Farchioni:2003bx}. Combined fits to $af_{xy}$ and $(am_{xy})^2$ from the two lightest ensembles with a mass cut of $am_{\rm avg}\leq0.01$ are shown in Fig.~\ref{fig:SU2fits}, whereas fitted parameters are included in Tab.~\ref{tab:fitres}. Here we would like to point out, that---in contrast to $\SU(2)\times\SU(2)$ $\chi$PT---in the partially quenched theory the same set of LECs (instead of a reduced set due to operator redundancies) appears as in $\SU(3)\times\SU(3)$, since we have to distinguish between sea and valence quarks. This distinction gives rise to a different functional dependence of the considered quantities on the sea and valence quark masses. (More correctly speaking, we use a $\SU(4|2)\times\SU(4|2)$ chiral Lagrangian and not a $\SU(2)\times\SU(2)$ one.)

Following this ansatz, as Fig.~\ref{fig:SU2fits} suggests, we did not
cure the problem of not being able to extend the fit range towards higher quark mass values. The important point is that our formulae do not contain any explicit dependence on the strange quark mass, whose physical value still lies outside the validity of the fit. The dependence on the background strange quark mass is implicitly contained in the LO and NLO fit parameters. One way (and in our opinion the most reliable one) to estimate this dependence would be to repeat the same analysis on a second set of ensembles, generated at a different dynamical value for $am_s$.

To compare these fit results with the previous ones obtained in $\SU(3)\times\SU(3)$ $\chi$PT (Sect.~\ref{subsec:su3Fits}), we use the formulae quoted in \cite{Gasser:1984gg} (cf. also \cite{Gasser:2007sg}) to match the three flavor $\chi$PT to the two flavor case at LO. The results for $aB_0$, $af_0$ and the low energy scales $\bar{l}_{3,4}$ (for a definition of the latter see \cite{Gasser:1984gg}) are shown in Tab.~\ref{tab:conv_SU3_SU2}. From the fact that the converted $\SU(3)\times\SU(3)$ fit results almost agree with the $\SU(2)\times\SU(2)$ fit results, one may argue that the effect of a slightly too high strange quark mass may be neglected for quantities in the pion sector. 

Turning the attention now towards the kaon sector, we will have to incorporate the strange quark mass value. Since we already saw that NLO-PQ$\chi$PT fails to describe our data in the region of the kaon mass or even beyond, we decided to demand chiral symmetry properties only for the two light quarks. Analogously to the heavy-light chiral perturbation theory in the B-sector \cite{Sharpe:1995qp, Booth:1994hx} we propose to use $\SU(2)\times\SU(2)$ $\chi$PT in the presence of $K$ mesons with terms of order $(m_\pi/m_K)^4$ being dropped at NLO. In other words, the $K$ mesons are now not treated as pseudo-Goldstone bosons.  
Under such considerations, we give, in the following, the fit formulae for the decay constant and squared mass of a meson made from a light valence quark with a mass $am_x$ and a heavier valence strange quark ($am_s$). Here the dynamical light quark mass ($am_l$) is taken into account as well, but the dynamical strange quark is viewed as a background field. (We followed the same ansatz to fit the kaon bag parameter $B_K$ \cite{Antonio:2007pb,AntonioCohen:Proc}.)
\begin{eqnarray}
\label{eq:SU2str_f}
af_K(\chi_x, \chi_l) &=& af_{0K}^{m_s}\,\bigg\{1\,+\,\frac{c_1^{m_s}}{(af_0)^2}\chi_l\,+\,\frac{c_2^{m_s}}{(af_0)^2}\chi_x\nonumber\\
&&\;-\,\frac1{(4\pi af_0)^2}\Big[\frac{\chi_x+\chi_l}{2}\log\frac{\chi_x+\chi_l}{2 (a\Lambda_\chi)^2}\,+\,\frac{\chi_l-2\chi_x}{4}\log\frac{\chi_x}{(a\Lambda_\chi)^2}\,\Big]\,\bigg\},       \\
\label{eq:SU2str_m}
\frac{(am_K)^2(\chi_x, \chi_l)}{\frac12 (am_s+am_{\rm res})} &=& 
aB_{0K}^{m_s}\,\bigg\{1\,+\,\frac{d_1^{m_s}}{(af_0)^2}\chi_l\,+\,\frac{d_2^{m_s}}{(af_0)^2}\chi_x\bigg\}.
\end{eqnarray}
Here the mass parameter $\chi_x \equiv 2aB_0(am_x+am_{\rm res})$ is used. The fit parameters $af_{0K}$, $aB_{0K}$, $c_{1,2}$, and $d_{1,2}$ all carry a superscript $m_s$ to indicate that these depend on the strange quark mass value. The parameters $af_0$ and $aB_0$ are the same as  the $\SU(2)\times\SU(2)$ $\chi$PT in the pure pion sector. Actually, in the following we fixed these to their values previously determined in the fits of the pure pion sector. 

\begin{figure}
\begin{center}
\hfill
\includegraphics[angle=-90, width=.45\textwidth]{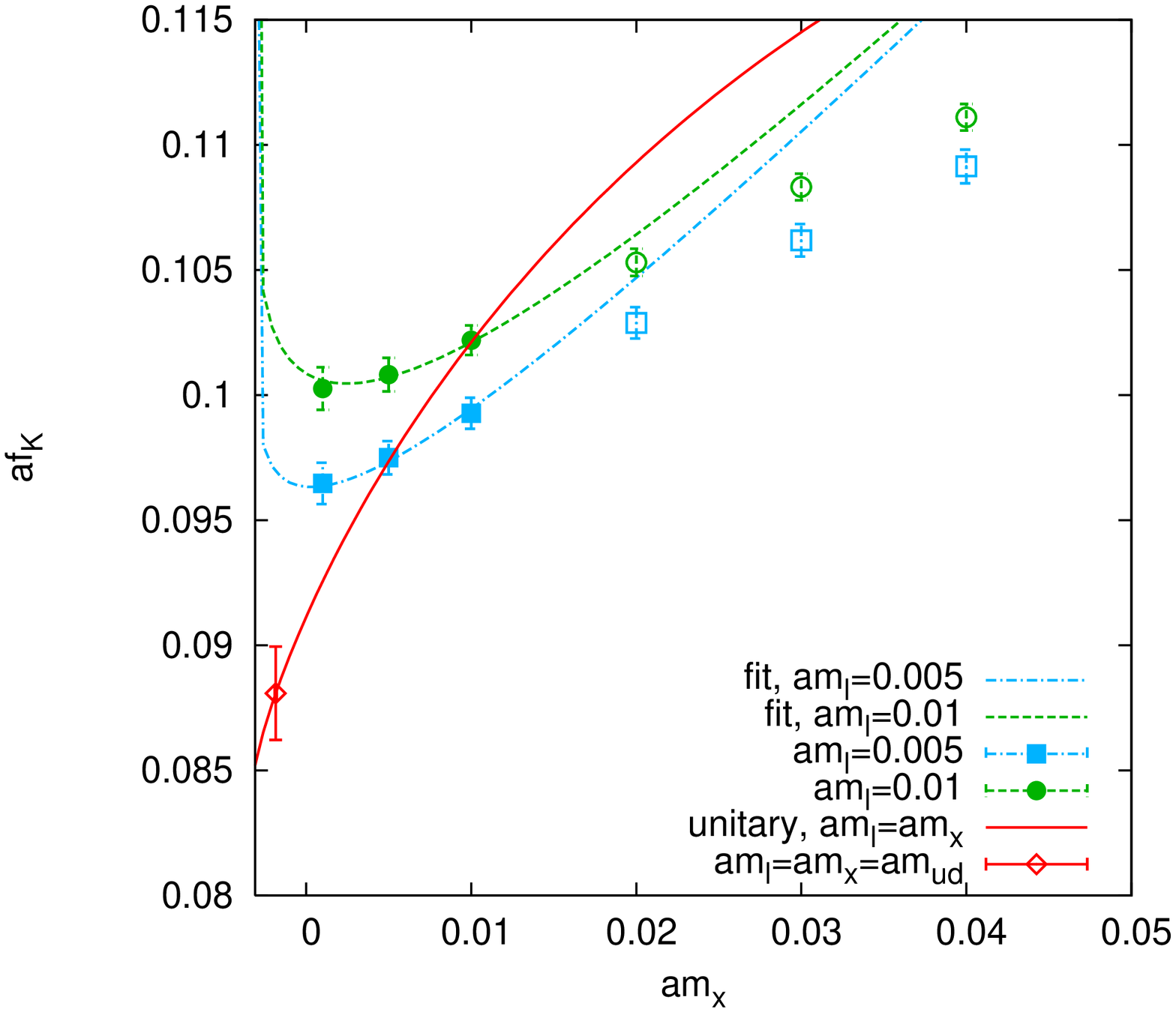}
\hfill
\includegraphics[angle=-90, width=.45\textwidth]{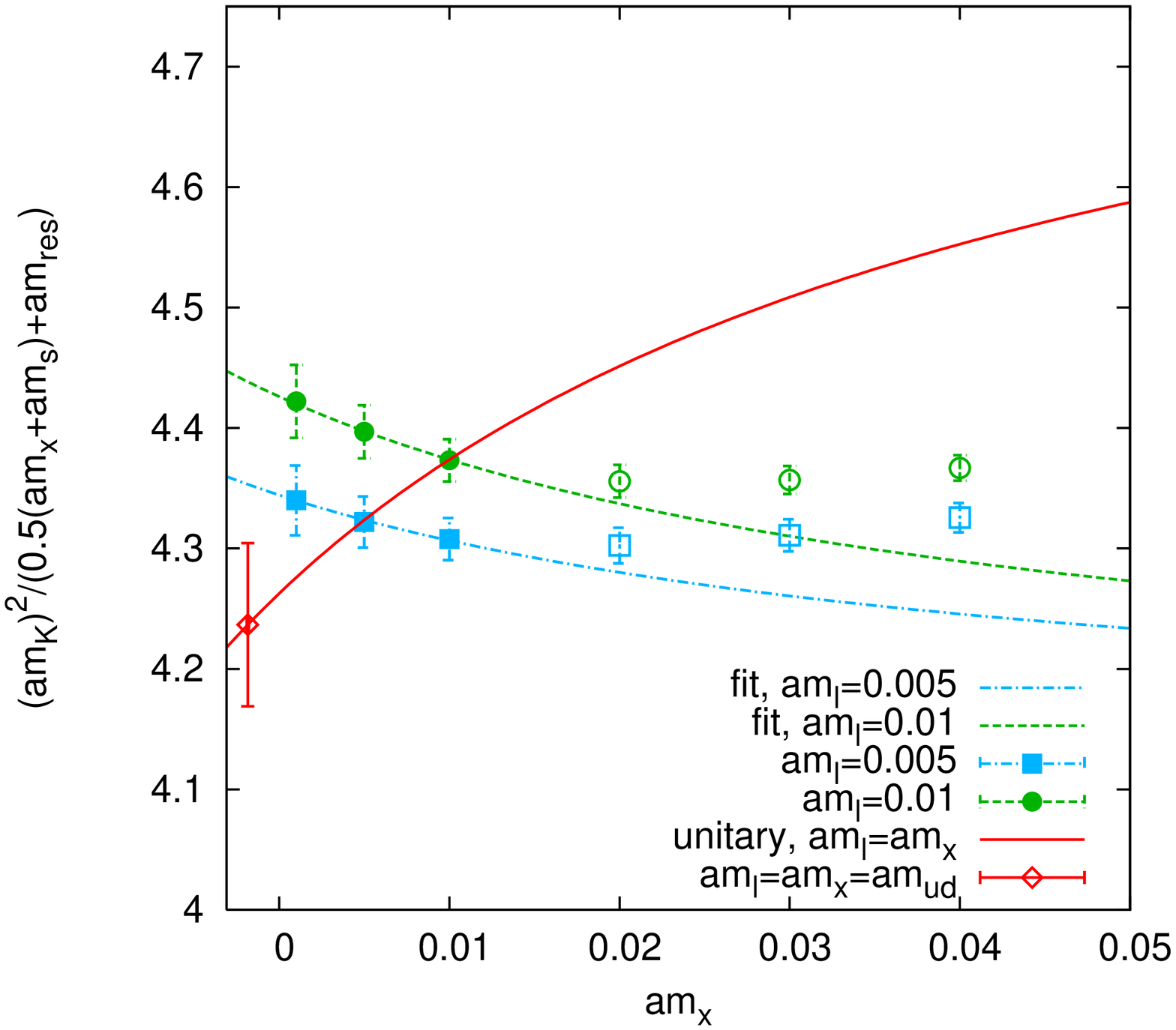}
\hfill
\end{center}
\vspace*{\closercaption}
\caption{\label{fig:SU2strange}$\SU(2)\times\SU(2)$ fits for the kaon sector. \textit{Left panel:} Kaon decay constant, \textit{right panel:} Kaon mass for $am_s=0.04$. Points with \textit{filled symbols} were included in the fit, while those with \textit{open symbols} were excluded.}
\vspace*{\afterFigure}
\end{figure}

We will use Eqs.~(\ref{eq:SU2str_f}, \ref{eq:SU2str_m}) to extrapolate the kaon decay constant and mass to the physical value of the light quark masses at a fixed value of the strange quark mass $am_s$. Repeating this for different values of $am_s$ allows us then to interpolate to the physical strange quark mass point as well. For the moment, since we only have data at one value for the dynamical strange quark mass, we can only vary the valence strange quark mass. For future runs, one should consider to have at least two sets of ensembles at different values of $am_s$ to allow for an interpolation between dynamical strange quark mass points. Finally, the fits were performed at $am_s=0.03$ and $0.04$, using all the points with light (dynamical, valence) quark masses $am_{l,x}\leq0.01$, i.e., the two ensembles with the lightest quark masses. Such fits at $am_s = 0.04$ are shown in Fig.~\ref{fig:SU2strange}, where the diamonds indicate the extrapolations to the physical light quark mass, $am_l^{\rm phys}$, at the fixed $am_s$.
By interpolating between the results at these two values for the strange quark mass, we are able to extract $af_K$ and $am_K$ at a determined physical strange quark mass, $am_s^{\rm phys}$, or, vice versa, use either the physical value (given the lattice spacing) of $f_K$ or $m_K$ to set $am_s^{\rm phys}$.

\begin{table}
\caption{\label{tab:conv_SU3_SU2}Comparison of converted $\SU(3)\times\SU(3)$ fit parameters with those from $\SU(2)\times\SU(2)$ fits. Low energy scales $\bar{l}_{3,4}$ are defined at $\Lambda=139\,{\rm MeV}$.}
\begin{center}
\begin{tabular}{l*{3}{c}c}\hline
                               &$aB_0$ & $af_0$ & $\bar{l}_3$ & $\bar{l}_4$ \\\hline 
$\SU(3)\times\SU(3)$, conv. & 2.457(78) & 0.0661(18) & 2.87(28) & 4.10(05) \\
$\SU(2)\times\SU(2)$ & 2.414(61) & 0.0665(21) & 3.13(33) & 4.43(14) \\\hline 
\end{tabular} 
\end{center}
\vspace*{\afterTable}
\end{table}

\section{\label{sec:physResults}Obtaining Physical Results}

First, we will discuss how the lattice scale was set and the points of physical quark masses were determined. In the remainder of this section we will utilize a non-perturbative renormalization scheme (RI/MOM) to obtain the quark masses in the $\overline{\rm MS}$ scheme at $\mu=2\,{\rm GeV}$.

\subsection{Determination of $am_l^{\rm phys}$, $am_s^{\rm phys}$, $a^{-1}$}

Given the reservations to use either the $\rho$-meson mass (width of
the resonance) or the Sommer-scale (ten percent uncertainty due to
phenomenological models) to set the lattice scale, here we will use
the mass of the  $\Omega^{-}$ baryon, a state made out of three
strange quarks, instead. One advantage of using this baryon mass, is
that up to NLO in $\chi$PT it is free of logarithms containing the
light quark masses~\cite{Tiburzi:2004rh}. Therefore the extrapolation of 
the measured masses
to the light physical mass can be easily performed using a linear ansatz without an uncertainty due to chiral logarithms. We used the measured $\Omega^{-}$ masses \cite{Lin:THESIS} with $am_s=0.03$ and 0.04 extrapolated to the light physical masses (using the configurations with $am_l=0.005$ and 0.01) to interpolate to the value of the physical strange quark mass.

The quark masses were obtained from the $\SU(2)\times\SU(2)$ fits described in Sect.~\ref{subsec:su2Fits}. For the light quark mass we solved for a pion mass of $m_\pi=135.0\,{\rm MeV}$, corresponding to the physical uncharged pion mass, while for the strange quark mass
the fit to the kaon mass was solved at $m_K=495.7\,\rm{MeV}$, which is the quadratically averaged neutral and charged kaon mass.

Since these two determinations depend on each other (the lattice scale is needed to convert the input masses into lattice units, whereas the quark masses are needed for the extrapolation in the light and interpolation in the strange quark masses for the baryon mass), we performed these two steps iteratively, starting with an initial guess for the quark masses. After eight iterations no further relevant change in the parameters were observed. The final values for $1/a$, $a$, $am_l$, $am_s$ can be found in Table~\ref{tab:results} (including only the statistical error).

Finally, with the knowledge of the values for the quark masses corresponding to their physical values, our chiral fits were used to extrapolate the meson decay constant to $f_\pi=124.1(3.6)\,{\rm MeV}$ and interpolate to $f_K=149.6(3.6)\,{\rm MeV}$ (statistical error only). Compared to their experimentally observed values \cite{Yao:2006px} of 130.7(0.1)(0.36) and 159.8(1.4)(0.44) MeV, our values are about five or six percent too low, but our measured ratio $(af_K)/(af_\pi)=1.205(18)$ agrees within the uncertainty with the experimental value of 1.223(12), indicating possible scaling effects in our results.

An interesting application of the latter result is to use it for the determination of the ratio $|V_{us}|/|V_{ud}|$ of CKM-matrix elements, as has been pointed  out in \cite{Marciano:2004uf}. Using the input for the branching ratios $\Gamma(K\rightarrow\mu\nu(\gamma))$ and $\Gamma(\pi\rightarrow\mu\nu(\gamma))$ plus radiative electroweak corrections from \cite{Yao:2006px}, we obtain $|V_{us}|/|V_{ud}|=0.2292(35)$ from our result for the decay constant ratio. This implies $|V_{us}|\;=\;0.2232(34)$, if $|V_{ud}|=0.97377(27)$ from super-allowed nuclear $\beta$-decays \cite{Yao:2006px} is taken into account. The quoted error combines both the errors from our determination of $f_K/f_\pi$ (statistical only) and the other input quantities. Here the main contribution comes from the decay constants, e.g., in the case of $|V_{us}|$ its contribution is 0.0033, whereas the other errors add up to 0.0005.  

\begin{table}
\caption{\label{tab:results}Determined lattice scale and spacing and unrenormalized quark masses ($am_x^{\rm phys}=am_x^{\rm bare}+am_{\rm res}$).}
\begin{center}
\begin{tabular}{*{6}{C}}\hline
  a^{-1}/{\rm GeV} & a/{\rm fm} & am_l^{\rm bare} & am_l^{\rm phys} & am_s^{\rm bare} & am_s^{\rm phys} \\\hline 
  1.729(28) & 0.1141(18)  & -0.001847(58) & 0.001300(58)  & 0.0343(16)   & 0.0375(16)\\\hline 
\end{tabular}
\end{center}
\vspace*{\afterTable}
\end{table}

\subsection{\label{subsec:npr}Non-Pertubative Renormalization and Quark Masses}

The renormalization factor $Z_m=1/Z_S$ needed to convert the extracted quark masses to the commonly used $\overline{\rm MS}$ scheme at a scale of 2 GeV has been calculated (amongst others) using $16^3\times32$, $L_s=16$ DWF configurations with $N_f=2+1$ flavors \cite{RBCUKQCD:npr}. (For details on the used configurations, cf.\ \cite{Allton:2007hx}.) We first matched the bare lattice operators to the RI/MOM scheme using the non-perturbative Rome-Southampton technique \cite{Martinelli:1994ty}, followed by a perturbative matching to the $\overline{\rm MS}$ scheme. Since DWF were used, we benefit from the controlled (small amount of) chiral symmetry breaking, resulting in $\mathcal{O}(a)$ improved operators/currents with reduced operator mixing.

In particular, we calculated the renormalization factor $Z_m$ in the \emph{regularization independent} (RI-)scheme according to
\begin{equation}
Z^{\rm RI}_m(ap)\;=\;\frac{Z_q}{Z_S}(ap)\,\frac{Z_A}{Z_q}(ap)\,\frac1{Z_A}\,,
\end{equation}
where the first two factors were obtained from the renormalized amputated vertex functions $\Lambda^{\rm ren}_S$ and $\Lambda^{\rm ren}_A$ ($\Lambda^{\rm ren}_{x}= (Z_x/Z_q) \Lambda_x$), respectively, and the last factor was obtained by measuring the appropriate hadronic matrix element (see \cite{Allton:2007hx}). The four loop matching from the RI to the \emph{renormalization group invariant} (RGI-)scheme has been applied to extract $Z_m^{\rm RI/MOM}(2\,{\rm GeV})$, which then was converted to the $\overline{\rm MS}$-scheme via three loop matching \cite{Chetyrkin:1999pq}. Finally, we get $Z_m^{\overline{\rm MS}}(2\,{\rm GeV})=1.656(48)(11)$, where the first error is the statistical one and the second one estimates the systematics due to residual chiral symmetry breaking. The latter was obtained from the difference which arises if instead of $\Lambda_A$ the combination $(\Lambda_A+\Lambda_V)/2$ is used in the determination of $Z_m$.

Using this result combined with the lattice spacing we obtain the quark masses via
\begin{equation}
m_x\;=\;Z_m^{\overline{\rm MS}}(2\,{\rm GeV})\,\cdot\,(1/a)\,\cdot\,am_x^{\rm phys}.
\end{equation}
The physical light quark mass (which, in fact, is the average up- and down-quark mass) we measure is $m_l=3.72(16)\,{\rm MeV}$, while for the strange quark mass we get a value of $m_s=107.3(4.5)\,{\rm MeV}$. (The quoted errors include the combined error from $Z_m^{\overline{\rm MS}}$ but only the statistical ones from other quantities.) This means we observe a quark mass ratio of $m_l:m_s\,=\,1:28.8(4)$.

\section*{Conclusions \& Outlook}
After realizing that fits to NLO $\chi$PT for three flavors are problematic up to the physical strange quark mass once sufficiently light quark masses have been reached, we found that using two flavor $\chi$PT for the pionic sector is a much more reliable approach. It eliminates the questionable dependence of the pion mass and decay constant on the strange quark mass value explicitly. Nevertheless, by converting our three flavor $\chi$PT fit parameters to the two flavor case, a sufficient agreement between the two approaches could be established. We quoted the fitted LO and NLO parameters for both the two and three flavor case.

By only demanding chiral symmetry properties for the two lightest quarks, we were able to apply $\SU(2)\times\SU(2)$ $\chi$PT to the kaon sector and successfully extracted the kaon mass and decay constant, despite the caveat that we had to include partially quenched strange quarks in that analysis because currently we are lacking data at a second value for the dynamical strange quark mass.

By using the experimentally measured values for $m_\pi$, $m_K$, and $m_{\Omega^-}$, we were able to extract the physical average light quark mass and strange quark mass, where for the conversion to the $\overline{\rm MS}(2\,{\rm GeV})$ via the RI/MOM scheme a non-perturbative renormalization technique was used. The pion and kaon decay constants were extrapolated or interpolated to these quark mass values. 
We also derived the ratio of CKM-matrix elements $|V_{us}|/|V_{ud}|$ from $f_K/f_\pi$. For the moment, no estimates for systematic errors (except for $Z_m$) are given, which we shall do in a forthcoming publication \cite{RBCUKQCD:24c}.

Currently, we are running simulations at a larger lattice volume ($32^3\times 64$, $L_s=16$), where also a second value for the strange quark mass will be included. (For a status report see \cite{Jung:Proc}.) These simulations will allow us to estimate the size of finite volume errors as well as to interpolate between dynamical strange quark mass values, resulting in a more reliable result for the kaon sector.

{\noindent\textbf{Acknowledgements.} We are thankful to all the members of the RBC and UKQCD-Collaborations. The computations were done on the QCDOC machines at University of Edinburgh, Columbia University, and Brookhaven National Laboratory. E.S.\ was supported by the U.S.\ Dept.\ of Energy under contract DE-AC02-98CH10886.}

\bibliography{references}

\bibliographystyle{JHEP-2} 

\end{document}